\documentclass[namedreferences]{SolarPhysics}
\usepackage[optionalrh]{spr-sola-addons} 
\usepackage{graphicx}        
\usepackage{color}           
\usepackage{url}             

\newcommand{\etal}{{\it et al.}}

\newcommand{\aap}{    {\it Astron. Astrophys.}}

\newcommand{\apj}{    {\it Astrophys. J.}}

\newcommand{\solphys}{{\it Solar Phys.}}

\begin{document}

\begin{article}

\begin{opening}

\title{Exact solutions for standing kink modes of the longitudinally stratified coronal loops }
\author{N. \surname{Dadashi$^{1}$}}
\author{H. \surname{Safari$^{2}$}}
\author{S. \surname{Nasiri$^{2}$}}
\author{Y. \surname{Sobouti$^{2}$}}
\runningauthor{Dadashi et al.} \runningtitle{Exact solutions of
standing kink modes } \institute{$^1$Zanjan University, Zanjan
45195, Iran,\\ $^2$Institute for Advanced Studies in Basic
Sciences, Gava Zang, Po. Box 45195-1159, Zanjan, Iran  }

\begin{abstract}

The influence of longitudinal structuring on the fast kink modes
of coronal loops is investigated. Analytical dispersion relations
and mode profiles are derived for the second-order ordinary
differential equation governing the z- component of the
perturbation in the magnetic field, $\delta B_z$. All other
components are given in terms of  $\delta B_z$. Deviations from
the frequencies and mode profiles of homogenous loops are given as
functions of the density scale height. The effects of the fixed
and variable column masses, negative scale heights, and density
contrasts inside and outside of the loops are studied. The
frequency ratios, mode profiles, and the antinode shifts from
those of the sine profiles of the homogenous loops, are tools to
estimate solar photospheric parameters. To this end, we have
expanded the relevant factors up to the second order in the
stratification parameter. In particular, we verify that the first
overtone antinode shifts are in the Mm range and are within the
reach of the resolutions of the present day observations.

\end{abstract}
\end{opening}

\section{Introduction}

 The first spatial oscillations of coronal loops were discovered
in extreme ultraviolet (171 $\AA$) by Aschwanden et al. (1999a).
Using a MHD wave theory developed by Edwin \& Roberts (1983),
Aschwanden et al. (1999b) and Nakariakov et al. (1999) interpreted
them as  the fundamental fast (kink) mode of  coronal loops.
Subsequently, Wang et al. (2003) observed frequency spectrum of
the standing slow modes of the loops.

Since then theoreticians have been supplementing the observed
data, oscillation periods, loop lengths, etc., with plausible
auxiliary parameters, such as plasma density, magnetic field
strength and configuration, etc., with the aim of obtaining
reasonably realistic models for the structure of the  loops
(Bennett et al., 1998; Verwichte et al., 2004; D\'{i}az et al.,
2002, 2004; Van Doorsselaere et al., 2004; Andries et al., 2005a,
b, Arregui et al., 2005; Dymova \& Ruderman, 2005; Erd\'{e}lyi \&
Fedun, 2006;  D\'{i}az \& Roberts, 2006; McEwan et al., 2006;
Donnelly et al., 2006;  Erd\'{e}lyi \& Verth 2007; Dymova \&
Ruderman, 2005; Dymova \& Ruderman, 2006; Verth et al., 2007
Doorsselaere et al., 2007; Erd\'{e}lyi \& Verth, 2007; Safari et
al., 2007; Karami \& Asvar 2007).

In their recent work, Dymova \& Ruderman (2005) and Safari et al.
(2007) reduce the MHD wave equations to a single Sturm-Liouville
equation for the z-component of the perturbation in the magnetic
field. Here, we use the formalism of Safari et al. and show that,
for an exponential plasma density stratification along the loop
axis, the problem has a closed analytical solution. The loop
model, equations of motion, and boundary conditions are presented
in Sec. 2. The closed solutions, including the dispersion
relation, are treated in Sec 3. Concluding remarks are given in
Sec. 4.

\section{Description of the model and equations of motion}
A coronal loop is approximated by a cylinder of length $L$ and
radius $R$. Loop ends are fixed at the photosphere. Loop curvature
is neglected, on account of $R<<L$. No initial flow is assumed
inside the loop. A uniform magnetic field along the axis pervades
the loop, $\mathbf{B}=B\hat{z}$. Gas pressure, gravity and all
dissipative and viscous forces are neglected. The density is
discontinuous on the lateral surface of the cylinder and varies
exponentially along the axis. Thus,
\begin{eqnarray}
\rho (r,z,\varepsilon) &=&
 \rho _i (\varepsilon )\exp(-\varepsilon z/L),\,\,~0\le z \le L/2,~{\rm inside ~tube},\nonumber \\
 &=&\rho _e (\varepsilon )\exp(-\varepsilon z/L),\,\,\,\,\,\hspace{2.2cm}{\rm outside ~tube},
 \end{eqnarray}
where $\varepsilon $ is the density scale height parameter, and
$\rho_i$, $\rho_e$ are the interior and exterior footpoint
densities, respectively. The assumption of exponential density is
in accord with the findings of Aschwanden et al. (1999). They
conclude this from their stereoscopic analysis of 30 loop
oscillations in EUV. Restriction of $z$ to the interval $[0-L/2]$
is permissible on account of the symmetry of the loop
configuration about its midpoint

The linearized ideal MHD equations are
\begin{eqnarray} &&\frac{{\partial \rho }} {{\partial t}} + \bf
\nabla .(\rho \delta \textbf{v}) = 0
,\,\,\,\,\,\,\,\,\,\,\,\,\,\,\,\,\,\,\,\,\,\,\,\,\,\,\,
{\rm Continuity ~Equation},\\
&& \frac{\partial\delta \textbf{v} } {{\partial t}} = \frac{1}
{{4\pi \rho }}(\bf \nabla  \times \delta \textbf{B}) \times
\textbf{B} ,\,\,\,\,\,\,{\rm Momentum ~Equation},
\\&&\frac{\partial\delta \textbf{B} } {{\partial t}} =
\bf \nabla \times (\delta \textbf{v} \times
\textbf{B}),\,\,\,\,\,\,\,\,\,\,\,\,\,\,\,\,\,\,\, {\rm Induction
~Equation }\
\\&&\bf \nabla .\,\delta \textbf{B}=0,\,\,\,\,\,\,\,\,\,\,\,\,\,\,\,\,\,\,\,\,\,\,\,\,\,\,\,\,\,\,\,\,\,\,\,\,\,\,\,\,\,\,\,\,\
{\rm Solenoidal ~Constraint},
\end{eqnarray}
where $\delta \textbf{v}$ and $\delta \textbf{B}$ are the Eulerian
perturbations in the velocity and magnetic fields, respectively.
An exponential $\varphi$ and $t$ dependence, is assumed,
$\exp[-i(m\varphi- \omega t)] $. By straightforward calculations
one can express all components of $\delta \bold{v}$ and $\delta
\bold{B}$ in terms of $\delta B_z$. The latter, in turn, is
obtained from the following second order PDE (See  Safari et al.
2007),
\begin{equation}
\left(\frac{{\partial ^2 }} {{\partial r^2 }} + \frac{1}
{r}\frac{\partial } {{\partial r}} + \frac{{\partial ^2 }}
{{\partial z^2 }} - \frac{{m^2 }} {{r^2 }} + \frac{{\omega ^2 }}
{{v_A^2 }}\right)\frac{\delta B_z}{B_0} = 0,\label{dbz}
\end{equation}
where $ v_A (z) = B/\sqrt {4\pi \rho (z)} $, local
Alfv$\acute{e}$n speed, has different values inside and outside of
the loop. Equation (\ref{dbz}), admits of a separable solution
$\delta B_z/B_0 = R(r)Z(z)$, where $R(r)$ satisfies Bessel's
equation and will not be further referred to here, and $Z$
satisfies the following
\begin{eqnarray}
&&\frac{d^{2}Z(x)}{dx^{2}}+\Omega^{2} e^{-\varepsilon x} Z(x)
=0,\,\,\,\
 0\leq x=z/L\leq 1/2,\label{z}\\
&&\Omega^2=\frac{L^2\omega^2}{v_{{A_i}}^2|_{\varepsilon=0}}
\frac{\tilde{\rho}_i(\varepsilon)+\tilde{\rho}_e(\varepsilon)}{2},\nonumber
\end{eqnarray}
where $\Omega$ and
$\tilde{\rho}_{i,e}=\rho_{i,e}(\varepsilon)/\rho_{i,e}(0)$  are
dimensionless frequency and footpoint densities, respectively.

    Equation (\ref{z}) is an eigenvalue
 problem weighted by $\exp{(-\varepsilon x)}$.
 Changing the variable $x$ to $\Omega\exp(-\varepsilon x/2)$
 reduces Eq.(7) to a Bessel
 equation with the following solutions
\begin{equation}
Z(x)=c_{1} J_{0}(2\frac{\Omega}{\varepsilon} e^{-\varepsilon x/2})
+ c_{2}Y_{0}(2\frac{\Omega}{\varepsilon} e^{-\varepsilon
x/2}),\label{zsol}
\end{equation}
where $c_1$ and $c_2$  are constants, and $J$ and $Y$ are Bessel
functions of first and second kind, respectively. The boundary
conditions are
\begin{eqnarray}
&&\label{bund1}{\rm~for\,\ odd\, modes} \left\{\begin{array}{cc}
  Z(0)=0,~~~~ \\
  Z'(1/2)=0,
\end{array}\right.\\
&&\label{bund2} {\rm~ for\, even\, modes} \left\{\begin{array}{cc}
  Z(0)=0,~~~ \\
  Z(1/2)=0.
\end{array}\right.
\end{eqnarray}\
Imposing the boundary condition at $z=0$ on Eq. (\ref{zsol}) gives
\begin{eqnarray}
c_1=Y_0 (2\frac{\Omega}{\varepsilon}),~~ c_2=-J_0
(\frac{2\Omega}{\varepsilon})
\end{eqnarray}\
 Substituting these coefficients in Eq. (\ref{zsol}) and imposing
 the boundary conditions at $x=1/2$ gives the dispersion relations
\begin{eqnarray}
 &&- Y_0 (2\frac{\Omega}{\varepsilon})J_1 (2\frac{\Omega}{\varepsilon} e^{ - \varepsilon /4} )+ J_0 (2\frac{\Omega}{\varepsilon})
 Y_1 (2\frac{\Omega}{\varepsilon} e^{ - \varepsilon /4} )=0,~
 ~{\rm odd ~modes},\nonumber\\\label{dispo} \\
 &&- Y_0 (2\frac{\Omega}{\varepsilon})J_0 (2\frac{\Omega}{\varepsilon} e^{ - \varepsilon /4})+
 J_0 (2\frac{\Omega}{\varepsilon})
 Y_0 (2\frac{\Omega}{\varepsilon} e^{ - \varepsilon /4}
 )=0,~~{\rm even ~modes}.\nonumber\\\label{dispe}
\end{eqnarray}

\begin{center}
\begin{figure}
\includegraphics{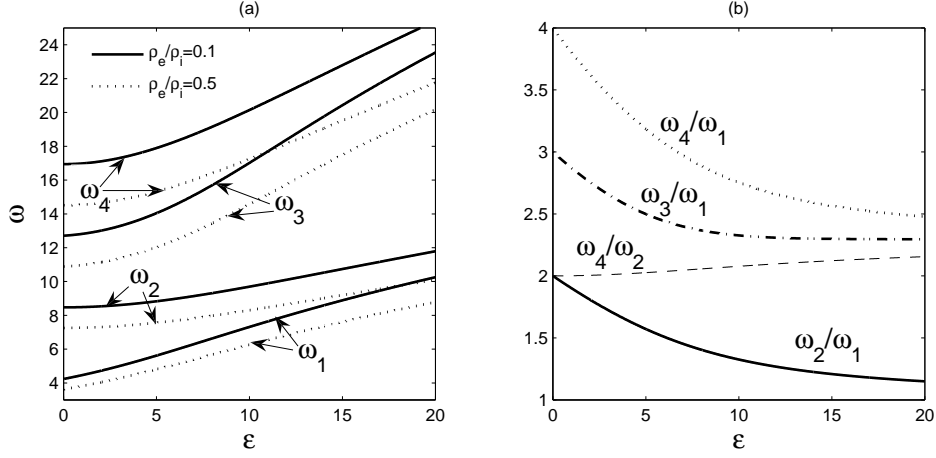}
 \vspace{6.8cm}
\caption[]{Fundamental and overtone frequencies  versus
$\varepsilon$. Solid lines for fixed column mass but variable
footpoint  densities, dashed lines for variable column masses but
fixed footpoint  density. All frequencies are in units of $\pi
v_{A_i}(\epsilon=0)/L$.}
           \label{fig1}
   \end{figure}
   \end{center}
Equations (\ref{dispo}) and (\ref{dispe}) are similar to those of
D\'{i}az \& Roberts (2006) with $W\to0$ in their analysis. In the
remainder of this section they are solved analytically for weakly
stratified loops and numerically for arbitrary stratifications.
\begin{center}
\begin{figure}
\includegraphics{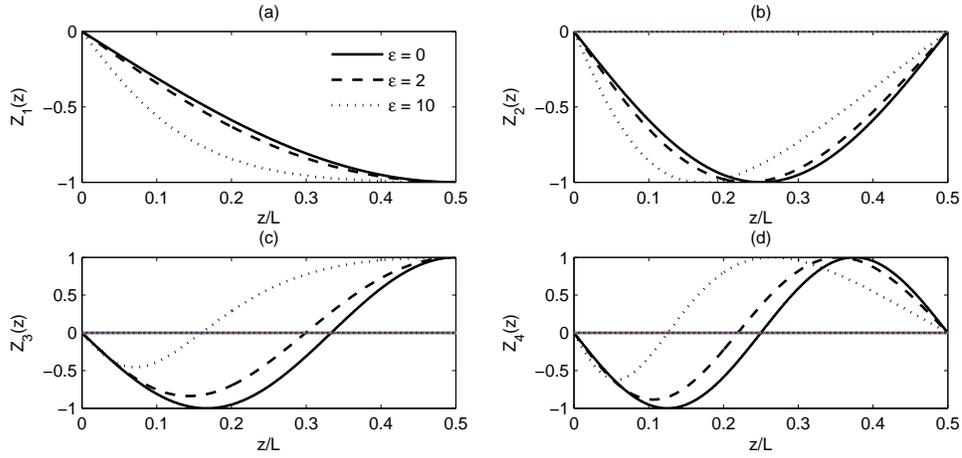}
      \vspace{6.cm}
\caption[]{ Mode profiles, $Z_n(z)$, \emph{a ,b , c, d},
corresponding to $n=1, 2, 3, 4$, respectively.  Solid, dashed, and
doted lines are for $\varepsilon=0$, $2$, and $10$, respectively.}
\label{fig2}
   \end{figure}  \end{center}

\subsubsection{Weak stratification}
Bessel functions can be expanded as
\begin{eqnarray}\label{bess}
   &&J_\nu(z)=\sqrt{\frac{2}{\pi
   z}}\left\{P_\nu(z)\cos z'-
   Q_\nu(z)\sin z'\right\},~z'=z-\left(\nu+\frac{1}{2}\right)\frac{\pi}{2},\nonumber\\
  &&Y_\nu(z)=\sqrt{\frac{2}{\pi z}}\left\{P_\nu(z)\sin z'+
   Q_\nu(z)\cos z'\right\},
\end{eqnarray}
where
\begin{eqnarray*}
P_\nu(z)\sim
1-\frac{(4\nu^2-1)(4\nu^2-9)}{128z^2}+O(\frac{1}{z^4}),~~Q_\nu(z)\sim\frac{(4\nu^2-1)}{8z}+O(\frac{1}{z^3}).\
\end{eqnarray*}

We insert Eq. (\ref{bess}) into  Eqs. (\ref{dispo}) and
(\ref{dispe}) and reduce them for $\varepsilon \ll1$ and find

\begin{equation}\label{weak}
\omega_n= \omega_1\left[\begin{array}{cc}
                 (2n-1)\left(1+\varepsilon(\frac{1}{2^2}+\frac{4}{(2n-1)^2\pi^2})+\varepsilon^2
                 (\frac{3}{2^7}+\frac{9}{2^2(2n-1)^2\pi^2})\right),&{\rm odd,} \\
              &\\   2n\left(1+\varepsilon\frac{1}{2^2}+\varepsilon^2(\frac{3}{2^7}-
                 \frac{1}{2^4n^2\pi^2}\right),~&{\rm even,} \\
               \end{array}\right.
\end{equation}
where $n=1,2,\cdots$, and $\omega_1=\frac{{\pi B}}{L}\left[2\pi
(\rho _i (0) + \rho _e (0))\right]^{-\frac{1}{2}}$ is the
fundamental kink frequency of homogenous loops. Expectedly,
$\omega_n\to n\omega_1$ as $\varepsilon\to0$. The following ratios
are noteworthy:
\begin{eqnarray}\begin{array}{cc}\label{ratio}
\frac{\omega_2}{2\omega_1}&\hspace{-2.5cm}=
1-\varepsilon\frac{101}{16\pi^2}
+\varepsilon^2\left(\frac{2525}{64\pi^4}+\frac{3535}{2048\pi^2}\right)
\\&\\
\frac{\omega_{2n-1}}{(2n-1)\omega_1}&\hspace{-.0cm}=1-\varepsilon\frac{25}{4\pi^2}
\frac{(2n-1)^2-1}{(2n-1)^2}+\varepsilon^2\frac{125}{2^4\pi^2}
(\frac{5}{\pi^2}+\frac{7}{2^5})\frac{(2n-1)^2-1}{(2n-1)^2} \\&\\
\frac{\omega_{2n}}{n\omega_2}&\hspace{-5.cm}=1+
\varepsilon^2\frac{1}{16\pi^2}\frac{n^2-1}{n^2}  \\
\end{array}
\end{eqnarray}
The frequencies and the ratio of any two odd numbered frequencies
begin decreasing linearly with $\varepsilon$. The ratio of two
even modes, however, begins decreasing quadratically with
$\varepsilon$. These features are also seen on the diagrams of Fig
\ref{fig3}$b$. Observational verification of these points,
however, has to await the availability of more extended and higher
resolutions data. Presently only two frequencies in three loops
are available ( Verwitche et al. 2004, Van Doorsselaere 2007).
\subsubsection{ \bf{Arbitrary stratification - Numerical approach}}
\begin{center}
\begin{figure}
\includegraphics{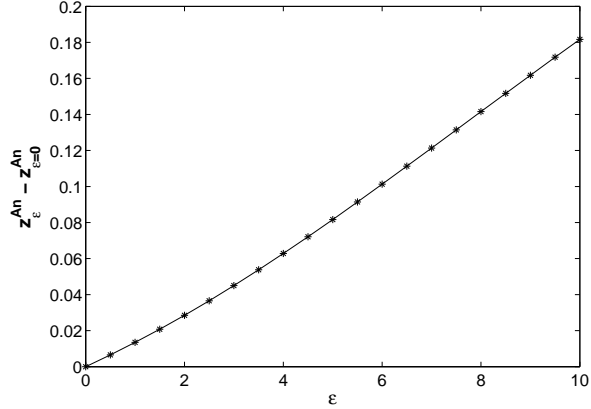}
      \vspace{5.5cm}
          \caption[]{Antinode shift, $z^{An}_\varepsilon -
          z^{An}_{\varepsilon=0}$,
          (normalized to $L$) against $\varepsilon$. The shift
           varies almost linearly with $\varepsilon$ }
           \label{fig3}
   \end{figure}
   \end{center}

We use Newton-Raphson's numerical method to solve Eqs.
(\ref{dispo}) and (\ref{dispe}) for the eigenfrequencies. In the
range, $0<\varepsilon=L/H<20$, the fundamental and three higher
kink frequencies, $\omega_{n}$,  $n=2,3$, and $4$, and the ratios,
$\omega_{n}/\omega_{1}$, are computed. The data are plotted in
Fig. \ref{fig3} for two density contrasts,
$\rho_e(\varepsilon)/\rho_i(\varepsilon)=0.1$ and 0.5. As the
density contrast increases the frequencies shift down. Their
ratios, however,  remain unchanged. The ratio $\omega_n/\omega_1$
begins with $n$ and decreases with increasing $\varepsilon$, in
compliance with Eqs. (\ref{ratio}).

Andries et al. (2005b) maintain that the frequency ratios could be
used as a seismological tool to estimate the coronal  density
scale heights. From the TRACE data, Verwichte et al. (2004) find
the ratio $\omega_2/\omega_1$ to be 1.64 and 1.81 for loops
designated by $D$ and $C$ in their records, respectively. Van
Doorsselaere, Nakariakov, \& Verwichte (2007) revisited the same
ratios from the observational data to be 1.58 and 1.82 for the
same loops, respectively, and 1.795 for another loop in their
current analysis. With the help of Fig. \ref{fig1}, we find the
corresponding $\varepsilon$ to be  4.98 and 1.9, respectively.
Assuming typical loop lengths, $L=100-400$ Mm, the density scale
heights fall in the range of $H=\varepsilon^{-1} L\simeq20-82$ and
$53-210$ Mm, respectively. These scale heights are  slightly
different from the findings of Andries et al. (2005a,b) and Safari
et al. (2007) for a sinusoidal density profile.

A noteworthy point is the effect of column mass on frequencies. In
Fig. \ref{fig1} we assume a constant
 footpoint density contrast, $\rho_e/\rho_i=0.1$,  and vary $\varepsilon$. Consequently, the
total column mass of the loop changes. Compared with variable
density contrast but fixed column mass, the mode profiles and the
frequency ratios remain unchanged. The frequencies themselves,
however, behave slightly differently. For variable column mass
models, the frequencies increase more sharply with $\varepsilon$.

The mode profiles, $Z(z)$ of Eq. (\ref{zsol}), are shown in Fig.
\ref{fig2} for $n=1,2,3$, and 4. For the unstratified case,
$\varepsilon=0$, the profiles are sinusoidal. With increasing
$\varepsilon$, they depart from the sine curves. Antinodes shift
toward the footpoints. Stronger the stratification, the greater
the shift is, in agreement with the findings of Safari et al.
(2007) and Verth et al. (2007). Van Doorsselaere et al. (2007)
point out, the shift of the antinodes is potentially, a coronal
seismological tool to estimate the density scale heights. In Fig.
\ref{fig3}, we have plotted the antinode shift,
$z^{An}_\varepsilon - z^{An}_{\varepsilon=0}$, of the first
overtone versus $\varepsilon$. The shift $\approx 0.02
\varepsilon$, grows approximately linearly with $\varepsilon$. Our
numerical result shows that, the shift in the antinode for
different  density contrasts , $\rho_e/\rho_i=0.1$  and $0.5$, are
the same. For typical loops, of lengths  $100 - 400$ Mm and
density scale heights, $H= 50$ and  $100$ Mm, the antinode shift
falls in the range $2.85\leq z^{An}_\varepsilon-
z^{An}_{\varepsilon=0}\leq 56.64$ Mm and $1.35 \leq
z^{An}_\varepsilon - z^{An}_{\varepsilon=0} \leq 25.12$ Mm,
respectively. Observation wise,   the resolution of current solar
satellite facilities, e.g., TRACE, SDO, SO, etc., seems adequate
to detect such antinode shifts and estimate the density scale
height of solar coronae. Verth et al. (2007) study semi circular
loops of sinusoidal density profiles and  find the antinode shifts
$\approx 0.028 \varepsilon$.
\begin{center}
\begin{figure}
\includegraphics{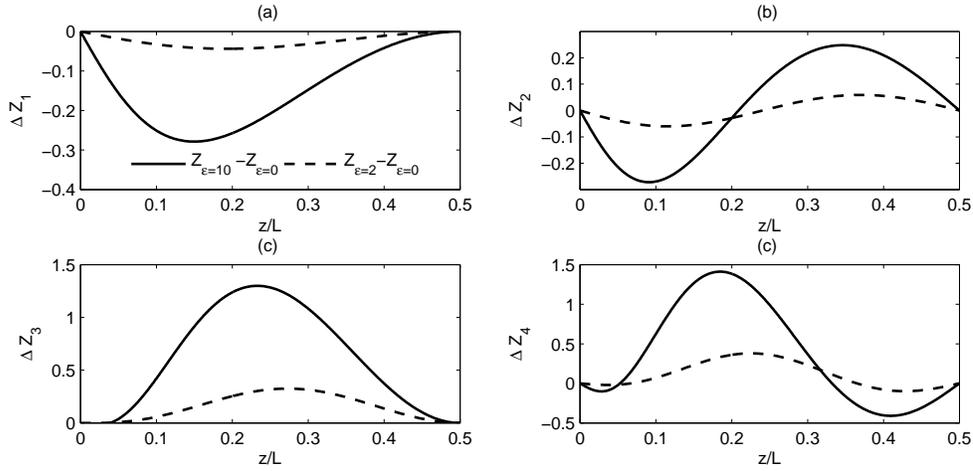}
      \vspace{6.cm}
          \caption[]{Differences between the eigenprofiles of the
          stratified and unstratified cases, $\Delta Z_l$, are plotted
z.}
    \label{fig4}
   \end{figure}
   \end{center}
\begin{center}
\begin{figure}
\includegraphics{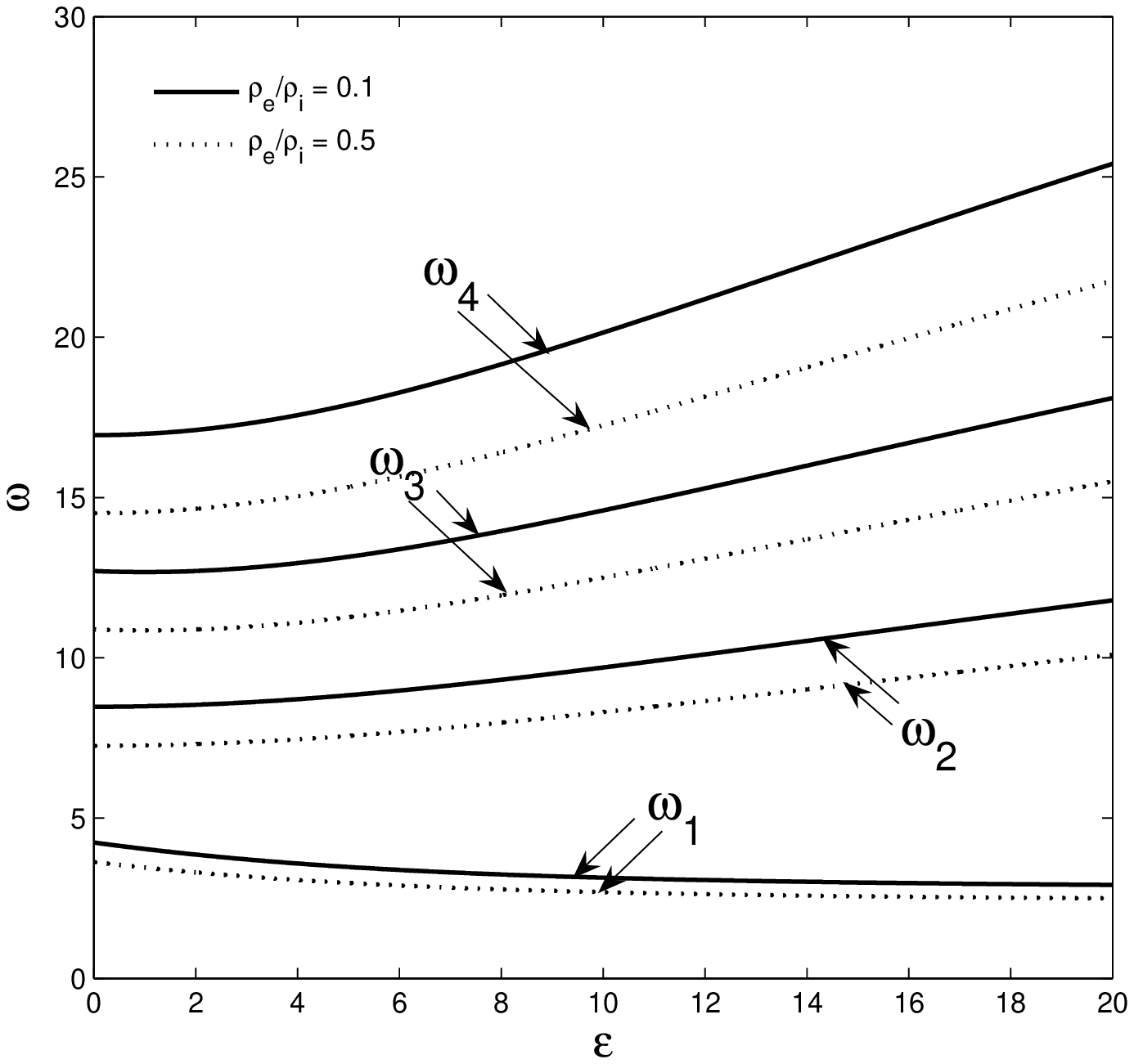}
      \vspace{5.5cm}
          \caption[]{ For the case of negative scale heights, the frequencies plotted
           versus $\varepsilon$, for two different density contrast,
           $\rho_{e}/\rho_{i}=0.1$ (solid lines) and $\rho_{e}/\rho_{i}=0.5$
           (dotted lines). }
           \label{fig5}
   \end{figure}
   \end{center}
\vspace{-2.25cm}Our numerical results show that, for a given
$\varepsilon$, the mode profiles are insensitive to changes in
density contrast, $\rho_e(\varepsilon)/\rho_i(\varepsilon)$. The
differences between the mode profiles of the stratified and
unstratified cases, $\Delta Z_n
=Z_n(\varepsilon,z)-Z_n(\varepsilon=0,z)$, $n=1,2,3,$ and 4 are
plotted in Fig. \ref{fig4}. Erd\'{e}lyi \& Verth (2007) maintains
that these differences in the mode profile are so small to be
resolved by current EUV instruments of TRACE mission.

Another interesting point is  negative scale heights, suggested by
Andries (2005a,b) on the basis of the error bars in the
observations of Verwichte et al. (2004). Here, the density at the
apex is higher than at  footpoints. Unlike the positive scale
scale height case: a) $\omega_1$ decreases with increasing
$|\varepsilon|$, see Fig. \ref{fig5}; b) higher overtones,
$\omega_n$, $n=2,3,\cdots$,  however, increase with
$|\varepsilon|$, though at slower rate; c) the ratios
$\omega_n/\omega_1$ increase  with $|\varepsilon|$ (not presented
in a diagram). As $|\varepsilon|$ increases, the mode profiles and
their node and antinodes move away from the footpoints and
concentrate more and more in the inner regions of the loop.
\section{Conclusions}
Suggested  theoretical models of 3D coronal loops are, still, far
from  the realities. Many complicating factors, such as variable
cross sections, variable magnetic fields, non-zero $\beta$
plasmas, etc., are to be accounted for in a realistic study of
both the equilibrium structure and the perturbed state of  actual
loops.  Here, we study the oscillations of loops with exponential
density variations along the loop axis.
\begin{description}
  \item[-] Analytical dispersion relations, Eqs. (\ref{dispo}) and (\ref{dispe}),
  and analytical mode profiles, Eq. (\ref{zsol}) and Fig. \ref{fig2}, are derived.
  \item[-] For weak stratifications, the kink frequencies and the frequency ratios
  are found up to the second order in $\varepsilon$, Eqs. (\ref{weak}) and (\ref{ratio}).
  \item[-]Increasing the density contrast decreases
 the frequencies but their ratios  and shape of the profiles remain unchanged.
\item[-] Models with  variable total column mass, but constant
footpoint densities,  are investigated. Compared with models of
constant total mass,  the frequencies  increase more sharply with
increasing  $\varepsilon$.
 \item[-] The case of negative scale heights is investigated and
 results are compared with those of positive $\varepsilon$'s.
   \item[-] For $1.58\leq \omega_2/\omega_1\leq 1.82$, and
  for typical loop lengths, 100-400 Mm,  the density scale heights fall in the
  range of 20-210 Mm, in agreement with
   Andries et al. (2005a, b),  Safari et al. (2007), McEwan et al.
   (2006), and Donnelly et al. (2006).
    \item[-] Based on our simple theoretical model and typical coronal  conditions,
     the antinode shift of the first overtone mode profiles are in the range of $1.3-56.6$
  Mm. They are in the range of the detectability of the resolution of the
  current observational instruments.

\end{description}

\end{article}
\end{document}